\documentclass[onecolumn,journal,draft,11pt]{IEEEtran}
\ifCLASSINFOpdf
\else
\fi
\DeclareMathAlphabet{\mathpzc}{OT1}{pzc}{m}{it}  %for Math Alphabets
\usepackage{mathrsfs}
\usepackage{amsthm}
\usepackage{makeidx}
\usepackage{examples}
\usepackage[pdftex]{graphicx}  
\usepackage{amssymb}
\usepackage{graphicx}
\usepackage{algorithm}
\usepackage{epstopdf}
\usepackage{color}
\usepackage{url}
\graphicspath{{../eps/}{../ps/}}
\usepackage{psfrag}    
\usepackage{algorithm}
\usepackage{algorithmic}
\usepackage{amsmath}
\usepackage[T1]{fontenc}
\usepackage[utf8]{inputenc}
\usepackage{tikz}
\usetikzlibrary{shapes,arrows}
\usepackage[numbers,sort&compress,square]{natbib}
\font\msbm=msbm10 at 10pt
\newcommand{\ZZ}{\mbox{\msbm Z}}
\newcommand{\RR}{\mbox{\msbm R}}

\newcommand{\FF}{\mbox{\msbm F}}
\def \Z {{\ZZ}}

\def \F {{\FF}}

\newtheorem{theorem}{Theorem}

\newtheorem{remark}[theorem]{Remark}
\newtheorem{example}[theorem]{Example}
\newtheorem{definition}[theorem]{Definition}

\definecolor{mypink1}{cmyk}{0, 0.7808, 0.4429, 0.1412}

\begin{document}
%\title{On Heterogeneous Regenerating Codes and Capacity of Distributed Storage Systems}
\title{On Optimal Heterogeneous Regenerating Codes}
%\title{Optimal Code on Heterogeneous Distributed Storage System}

%\author{Author~1 and~Author~2}

%\author{Krishna~Gopal~Benerjee and~Manish~K.~Gupta,~\IEEEmembership{Senior~Member,~IEEE}\thanks{The authors are with Dhirubhai Ambani Institute of Information and Communication Technology Gandhinagar, Gujarat, 382007 India (email: krishna\_gopal@daiict.ac.in, mankg@computer.org)}}

%\author{
%\IEEEauthorblockN{Author~1~and~Author~2} }

\author{
\IEEEauthorblockN{Krishna Gopal Benerjee and Manish~K.~Gupta,~\IEEEmembership{Senior~Member,~IEEE}}\\
\IEEEauthorblockA{Dhirubhai Ambani Institute of Information and Communication Technology Gandhinagar, Gujarat, 382007 India\\
Email: krishna\_gopal@daiict.ac.in, mankg@computer.org}
%\and
%\IEEEauthorblockN{Manish K. Gupta}
%\IEEEauthorblockA{Laboratory of Natural Information Processing\\Dhirubhai Ambani Institute of Information\\and Communication Technology\\Gandhinagar, Gujarat, 382007 India\\
%Email: mankg@computer.org}
%

}


%Telephone: (800) 555--1212, Fax: (888) 555--1212}


%\markboth{IEEE COMMUNICATIONS LETTERS,~Vol.~1, No.~1, February~2016}{Benerjee and Gupta: On Optimal Heterogeneous Regenerating Codes}

\maketitle

\begin{abstract}
Heterogeneous Distributed Storage Systems (DSSs) are close to the real world applications for data storage. Each node of the considered DSS, may store different number of packets and each having different repair bandwidth with uniform repair traffic. For such heterogeneous DSS, a failed node can be repaired with the help of some specific nodes. In this work, a family of codes based on graph theory, is constructed which achieves the fundamental bound on file size for the particular heterogeneous DSS. %The tradeoff curve between storage and repair bandwidth is studied for such heterogeneous DSS. By analyzing the capacity formula new minimum bandwidth regenerating (MBR) and minimum storage regenerating (MBR) points are obtained on the curve. It is shown that in some cases, these are better than the homogeneous DSS.
%\par Distributed Replication-based Simple Storage Codes consists of an inner Fractional Repetition (FR) code and an outer MDS code. FR code is smart distribution of encoded replication information packets on nodes in $(n, k)$ heterogeneous distributed storage system with dynamic replication factor. On the basis of node-packet distribution incidence matrix, the FR code is classified. In practical scenario it is desirable to have more popular nodes having higher storage capacity. Motivated by this we propose a new framework called flower code to design FR code for high parameters such that some flower codes are universally good.

\end{abstract}


% \fi

%\begin{IEEEkeywords} Heterogeneous Distributed Storage System, Graph Construction, Minimum Bandwidth Regenerating Cods \end{IEEEkeywords}
\IEEEpeerreviewmaketitle

%%%%%%%%%%%%%%%%%%%%%%%%%%%%
\section{Introduction}
%\footnote{Dhirubhai Ambani Institute of Information and Communication Technology Gandhinagar, Gujarat, 382007 India\\	Email: krishna\_gopal@daiict.ac.in, mankg@computer.org}
\IEEEPARstart DATA storage is a big challenge for mankind since ancient times. Recently emerged Cloud computing provides an excellent way to store the data in a Distributed Storage Systems (DSSs). %Many such commercial systems are in use such as Hadoop based DSS of Facebook, Yahoo,  IBM, Amazon and Microsoft Windows Azure system \cite{XorbasVLDB,Huang:2012:ECW:2342821.2342823,skydrive,amazonec2}. 
In such a DSS, data file is stored on $n$ distinct nodes in such a smart way that the complete file can be retrieved by connecting certain number of nodes. %Hence data reliability is a major challenge for researchers.  %For example, in one month the maximum number of node failures is approximately 110 out of 3000 nodes in Facebook clusters \cite{XorbasVLDB}. 
 In the case of node failure, system has to repair the failed node by either generating functional equivalent of the data loss or by generating the exact data that was lost on that node. In order to provide reliability systems use either simple replication or MDS (maximum distance separable) erasure codes. Simple replication uses more space (so it is bad for storage minimization) and erasure MDS code approach is not efficient for bandwidth minimization in a node repair process. To optimize these conflicting parameters data storage and bandwidth, in a seminal paper Dimakis et. al \cite{dgwr7} introduced family of codes called regenerating codes. These regenerating codes received attention in several papers \cite{dgwr7, survey,dress11,DBLP:journals/corr/abs-1302-3681}.

Consider a DSS of total $n$ distinct nodes. Homogeneous Regenerating codes are specified by the parameters $[n, k, d,\alpha,\beta, B]$, where $B$ is the file size and $\alpha$ is the number of packets on each node. In order to get a file, user has to connect $k$ ($< n$) nodes out of total $n$ nodes. The particular $\alpha$ and $k$ are known as node storage capacity and reconstruction degree for the DSS ~\cite{5550492}. In case of a node failure, data can be recovered by contacting $d$ nodes and downloading $\beta$ packets from each node. Thus total bandwidth for a repairing a node is $d\beta$, where $d$ and $\beta$ are known as repair degree and repair traffic respectively. By optimizing both $\alpha$ and $\beta$ in different order, we get two kind of regenerating codes called  Minimum Storage Regenerating (MSR) codes and Minimum Bandwidth Regenerating (MBR) codes. %MSR codes useful for archival purpose and MBR codes useful for Internet applications. %Some of the MBR codes studied by the researchers fails to optimize other parameters of the system such as disk I/O, computation and scalability etc. Towards this goal a class of MBR codes called Dress codes were introduced and studied by researchers \cite{rr10} to optimize disk I/O. These codes have a repair  mechanism known as encoded repair or table based repair. Dress codes consisting of an inner code called fractional repetition (FR) code  and outer MDS code. Construction of FR codes has been an important research problem and many constructions of FR codes are known based on bipartite graph \cite{DBLP:journals/corr/abs-1102-3493}, resolvable designs \cite{DBLP:journals/corr/abs-1210-2110}, regular graphs \cite{rr10,Wangwang12} and other \cite{Gupta:arXiv1302.3681}. %%%cite latest gupta srijan aaron paper
Many researchers constructed MBR and MSR codes using combinatorial designs, graphs, sequences and matrices etc \cite{7273884,KGBenerjee}.

%%%%%problem with this paper
For ($n,k$) heterogeneous DSS with dynamic node storage capacity, repair traffic, repair degree and reconstruction degree, a fundamental bound is established in \cite{DBLP:journals/corr/BenerjeeG15}. The computational complexity to calculate parameters, for the fundamental bound achieving codes, is very high. Hence, in this work, we consider a special case (considered in \cite{DBLP:journals/corr/BenerjeeG14}) by choosing constant repair traffic and reconstruction degree. Further, we calculate relations on parameters of codes which achieves the bound. For the particular parameters, the optimal codes (the fundamental bound achieving codes) are constructed using graphs.

\par \textit{Organization}: The paper is organized as follows. Section $2$ describes the model of heterogeneous DSS and collects the necessary background including fundamental bound on file size for the heterogeneous DSS. %is computed in Section $3$. Condition for the codes achieving the $min$-$cut$ bound is analyzed in Section $4$. 
Conditions for the optimal codes are established in Section $3$. Graphical construction of the family of such optimal codes is given in Section $4$. % and analysis of Theorem $6$ is given in Section $5$. 
Final section concludes the paper with general remarks.
 %a brief overview of heterogeneous distributed storage system which is considered in the whole paper is provided in next section. Some worked analysis related to capacity of heterogeneous DSS, are in section 3. Section 4 is all about the proof and explanation of the results. Finally we conclude the paper with reference.   %classification of FR code (on $(n, k)$ heterogeneous DSS) having dynamic replication factor of packets. The third section is based on construction of FR code called \textit{flower} code. Some of the flower codes are universally good. Finally we conclude the paper with appendix. 
%In this paper, we present a simple and elegant construction of FR codes known as Flower codes. The construction gives us both strong and FR codes.The rest of the paper is organized as follows. In Section $2$ we collect basics of Fractional Repetition Codes. In Section $3$, 
%we present an algorithm  for constructing $n \times \theta$ incidence matrix of node-packet distribution of FRC.  In Section $4$ we present  algorithms for constructing regular graph and hence Fractional Repetition codes for $n=\theta$.   Finally Section  $5$ concludes the paper with general remarks.

 
%\hfill January 11, 2007
%%%%%%%%%%%%%%%%%%%%%%%%
\section{Model} 
In heterogeneous DSS, a file is divided into encoded packets and the encoded packets are distributed among $n$ distinct nodes $U_i$ ($i = 1,2,\ldots, n$) such that each node has storage capacity $\alpha_i$ and repair degree $d_i$.
An user can reconstruct the file by downloading data from any $k$ $(< n)$ nodes.  If a node $U_i$ fails then data collector will download $\beta$ packets from specific $d_i$ nodes out of remaining $n-1$ nodes. The particular $d_i$ nodes are called helper nodes for the failed node $U_i$. In such a case, repair bandwidth for a node $U_i$ is $\gamma_i = d_i\beta$. 

An example of such heterogeneous DSS is illustrated in Figure \ref{DSS example}. In this example, a file with size $3$ $(=B)$ is stored in ($n=6,k=2$) heterogeneous DSS with repair traffic $\beta$ is $1$. In the particular DSS, node storage capacity $\alpha_i$ is $2,2,2,3,2$ and $2$ for $i=1,2,3,4,5,6$ (see Figure \ref{DSS example}). Note that $\alpha_i$ = $\gamma_i$ = $d_i$ for each $i= 1,2,3,4,5,6$. %In this example, a file with size $3$ (= $B$) is stored on $6$ (= $n$) distinct nodes in distributed fashion with reconstruction degree $2$ (= $k$). In the particular DSS, node storage capacity $\alpha_i$ and repair degree $d_i$ are not identical for each node $U_i$ $(1\leq i\leq 6)$. A failed node $U_i$ can be repaired by downloading $1$ $(=\beta)$ packet from each node of a set having specific $d_i$ $(d_4=3$ and $d_i=2 \mbox{ for }i=1,2,3,5,6)$ helper nodes. The set of helper nodes is the specific subset of remaining $5$ $(=n-1)$ node. For $i=1,2,3,5,6$, node repair bandwidth $\gamma_i$ is 2 and $\gamma_4=3$.
%\begin{remark}
%In this paper, single node failure is considered since multi-node failure can be assumed as a sequence of single node failure within a small time interval.
%\end{remark}
%\begin{remark}
	%In the heterogeneous DSS, node repair bandwidth $\gamma_i$ depends on repair degree $d_i$ and repair traffic $\beta$. By varying any one parameter between repair degree $d_i$ and repair traffic $\beta$, one can find DSS with dynamic node repair bandwidth $\gamma_i$. Thus, in this DSS model $\beta$ is taken a constant for each node.
%	For ($n,k$) heterogeneous DSS with dynamic node storage capacity, repair traffic, repair degree and reconstruction degree, the min-cut bound is established in \cite{DBLP:journals/corr/BenerjeeG15}. The computation complexity to calculate parameters for the min-cut bound achieving codes, is very high. Hence, in this work, we consider a special case (considered in \cite{DBLP:journals/corr/BenerjeeG14}) by taking repair traffic as constant and calculate some parameters for some bound achieving codes.
%\end{remark}
%For a reliable DSS, a failed node $U_i$ $(1\leq i\leq n)$ is repaired with the help of some specific $d_i$ helper nodes. Set of such helper nodes can be called surviving set. Formally, 

Again, a set of $d_i$ helper nodes which are used for repairing the failed node $U_i$, is called as \textit{surviving set}. Formally, surviving sets are defined as follows.
% Note that there could be several surviving sets for a given node $U_i$. Indexing all the distinct surviving sets by a positive integer $\ell_i,$ let us denote them by $S_i^{(\ell_i)}$ ($\ell_i=1,2,\ldots$). For a particular node $U_i$, number of distinct surviving sets are finite say  $\tau_i$ then $\ell_i=1,2,...,\tau_i$.
\begin{definition}(Surviving Set): In a $(n,k)$ heterogeneous DSS, surviving set of a node $U_i\;(1\leq i\leq n)$ is a set of $d_i$ nodes which are used for repairing the node $U_i$. Note that there could be several surviving sets for a given node $U_i$. Indexing all the distinct surviving sets by a positive integer $\ell_i,$ let us denote them by $S_i^{(\ell_i)}$ ($\ell_i=1,2,\ldots$). For a particular node $U_i$, number of distinct surviving sets are finite say  $\tau_i$ then $\ell_i=1,2,...,\tau_i$.
\end{definition}
%For the particular node $U_4$, one can find surviving sets $ S_4^{(1)} =  \{U_1, U_3, U_5\},$ $ S_4^{(2)} =  \{U_2, U_3, U_5\}$, $ S_4^{(3)} = \{U_1, U_6, U_5\},$ and $S_4^{(4)} = \{U_2, U_6, U_5\}$ (see Figure \ref{DSS example}). Note that $\tau_i$ = $2,4,4,4,2,2$ for $i=1,2,3,4,5,6$.

%\begin{figure}
	%\centering
	%\includegraphics[scale=0.3]{flowerfigures/example}

		% Define block styles
		\tikzstyle{line} = [draw, -]
		\tikzstyle{cloud} = [draw, circle,fill=red!20, node distance=2cm, minimum height=2em]
		\tikzstyle{block 1} = [rectangle, draw, fill=red!20, text width=8em, text centered, rounded corners, minimum height=1em, node distance=3.5cm]
		\tikzstyle{block 2} = [rectangle, draw, fill=red!20, text width=8em, text centered, rounded corners, minimum height=1.5em, node distance=1cm]
		\tikzstyle{block 3} = [rectangle, draw, fill=red!20, text width=10em, text centered, rounded corners, minimum height=1.5em, node distance=1cm]
		\tikzstyle{block 4} = [rectangle, draw, fill=red!20, text width=8em, text centered, rounded corners, minimum height=1.5em, node distance=4.8cm]
		%\tikzset{vertex/.style = {shape=circle,draw,minimum size=0em}}
		\tikzset{edge/.style = {->,> = latex'}}
		\begin{figure}\center
			\begin{tikzpicture}[node distance = 2cm, auto]
			% Place nodes
			%\node [cloud] (1) {$v_1$};
			%\node [cloud, left of=1] (2) {$v_2$};
			%\node [cloud, below of=1] (identify) {$v_4$};
			%\node [cloud, left of=identify] (evaluate) {$v_3$};
			%\path [line] (1) -- (identify);
			%\path [line] (identify) -- (evaluate);
			%\path [line] (1) -- (evaluate);
			%\path [line] (2) -- (identify);
			%\path [line] (2) -- (evaluate); 
			%\draw (0.7,-0.5) node {$y_1$, $y_2$};
			%\draw (-2.7,-0.5) node {$y_3$, $y_4$};
			%\draw (-2.7,-1.5) node {$y_5$, $y_6$,};
			%\draw (-2.7,-2) node {$y_7$};
			%%\draw (-4.5,-3.5) node {$x_1+x_3$,};
			%%\draw (-5.2,-4) node {$x_1+x_2+x_3+x_4$};
			%\draw (0.7,-1.5) node {$y_8$, $y_9$,};
			%\draw (0.7,-2) node {$y_{10}$};
			%\draw (1.5,-1) node {$\Longleftrightarrow$};
			\draw (-2.2,0) node {$U_1$};
			\draw (-2.2,-1) node {$U_2$};
			\draw (-2.2,-2) node {$U_3$};
			\draw (-2.2,-3) node {$U_4$};
			\draw (-2.2,-4) node {$U_5$};
			\draw (-2.2,-5) node {$U_6$};
			
			\draw (0,-0.4) node {$\alpha_1=2$};
			\draw (0,-1.45) node {$\alpha_2=2$};
			\draw (0,-2.45) node {$\alpha_3=2$};
			\draw (0,-3.45) node {$\alpha_4=3$};
			\draw (0,-4.45) node {$\alpha_5=2$};
			\draw (0,-5.45) node {$\alpha_6=2$};
			%\draw (-1.5,-0.4) node (n1) {};
			%	\draw (1.5,-0.4) node (n2) {};
			%\path [line] (n1) -- (n2);
			
			%	\node[vertex] (a) at  (0,0) {};
			%	\node[vertex] (b) at  (4,3) {};
			\draw[edge] (0.6,-0.4) to (1.5,-0.4);
			\draw[edge] (-0.6,-0.4) to (-1.5,-0.4);
			
			\draw[edge] (0.6,-1.4) to (1.5,-1.4);
			\draw[edge] (-0.6,-1.4) to (-1.5,-1.4);
			
			\draw[edge] (0.6,-2.4) to (1.5,-2.4);
			\draw[edge] (-0.6,-2.4) to (-1.5,-2.4);
			
			\draw[edge] (0.6,-3.4) to (1.85,-3.4);
			\draw[edge] (-0.6,-3.4) to (-1.85,-3.4);
			
			\draw[edge] (0.6,-4.4) to (1.5,-4.4);
			\draw[edge] (-0.6,-4.4) to (-1.5,-4.4);
			
			\draw[edge] (0.6,-5.4) to (1.5,-5.4);
			\draw[edge] (-0.6,-5.4) to (-1.5,-5.4);
			
			\node [block 1] (3) {$x_1$, $x_1+x_2+x_3$};
			\node [block 2, below of=3] (4) {$x_1$, $x_2$};
			\node [block 2, below of=4] (5) {$x_2$, $x_3$};
			\node [block 3, below of=5] (6) {$x_1$, $x_3$, $x_1+x_3$};
			\node [block 2, below of=6] (7) {$x_2$, $x_1+x_3$};
			\node [block 2, below of=7] (8) {$x_3$, $x_1+x_2+x_3$};
			
			\draw (1.9,-5.1) node {\textcolor{mypink1}{{\Huge $\times$} failure}};
			\draw[edge] (1.52,0) to (3.6,-4.7);
			\draw[edge] (1.85,-3.2) to (3.5,-4.7);
			\draw (3.4,-3) node {$\beta=1$};
			\draw (2.5,-4.3) node {$\beta=1$};
			\node [block 4, right of =8] (empty) {new node $U_6$};
			%\draw (4.7,-5.5) node {new node};
			
			\draw (4,-1) node {File: ($x_1,x_2,x_3$)};
			\draw (4,-1.5) node {($B=3$)};
			
			\end{tikzpicture}
	%\caption{A file is divided into $3$ distinct coded packets $x_1,$ $x_2$ and $x_3$ from field $\F_q$. These three packets are encoded into five distinct packets and some copies of the five packets are distributed among $6$ nodes such a way that any data collector can download whole file by contacting any $2$ nodes. In this ($6,2$) heterogeneous DSS, repair degrees and the number of packets on each storage node $U_i\;\ (1\leq i\leq 6)$ are $2,2,2,3,2,2$ respectively. To repair a failed node $U_i$, each helper node of a surviving set $S^{(\ell_i)}_i$ (for $\ell_i\in\{1,2,\ldots,\tau_i\}$) will download $1$ packet to repair the  node $U_i\ i.e.\ \beta = 1$.}
%\caption{A file is divided into $3$ ($=B$) distinct coded packets $x_1,$ $x_2$ and $x_3$ on field $\F_q$. These three packets are encoded into thirteen distinct packets. The thirteen encoded packets are distributed among $6$ ($=n$) distinct nodes such that any data collector can download whole file by contacting any $2$ ($=k$) nodes. In the ($6,2$) heterogeneous DSS, node storage capacity $\alpha_i$ is $2,2,2,3,2$ and $2$ for $i=1,2,3,4,5,6$. Here node storage capacity and repair degree are identical $i.e.$ $\alpha_i$ = $d_i$ for each $i$. Observe that repair traffic $\beta$ is $1$ for each helper node of arbitrary node failure $U_i$.}
\caption{A file is divided into $3$ ($=B$) distinct coded packets $x_1,$ $x_2$ and $x_3$ on field $\F_q$. These three packets are encoded into thirteen distinct packets and distributed in ($6,2$) heterogeneous DSS.}
	\label{DSS example}
\end{figure}
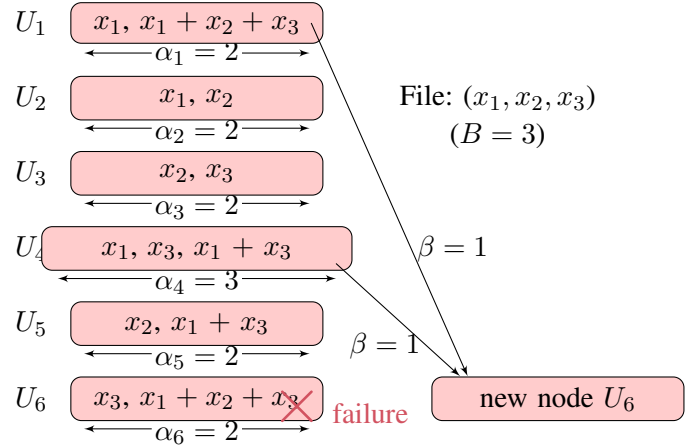

For $(6, 2)$ heterogeneous DSS (as shown in Figure \ref{DSS example}), the surviving sets of each node are listed in Table \ref{example table}.
\begin{table}[ht]
	\caption{Surviving sets for nodes of $(6, 2)$ heterogeneous DSS considered in Figure \ref{DSS example}.}
	\centering 
	\begin{tabular}{|c||c|c|}
		\hline
		Nodes&All possible surviving sets& \# surviving sets\\[0.5ex]
		$U_i$& $S_i^{(\ell_i)}$                                          &   $\tau_i$    \\
		\hline\hline
		$U_1$&$S_1^{(1)} = \{U_4, U_6\}, S_1^{(2)} = \{U_2, U_6\}.$      &    $2$  \\
		\hline
		$U_2$&$S_2^{(1)} = \{U_1, U_3\}, S_2^{(2)} =  \{U_1, U_5\},$     &    $4$   \\
		&$ S_2^{(3)} = \{U_4, U_3\},S_2^{(4)} =  \{U_4, U_5\}$.     &        \\
		\hline 
		$U_3$&$S_3^{(1)} = \{U_2, U_4\}, S_3^{(2)} =  \{U_2, U_6\},$      &    $4$  \\
		&$ S_3^{(3)} = \{U_5, U_4\},S_3^{(4)} =  \{U_5, U_6\}$.      &     \\
		\hline 
		$U_4$&$ S_4^{(1)} =  \{U_1, U_3, U_5\}, S_4^{(2)} =  \{U_2, U_3, U_5\},$&$4$\\
		&$ S_4^{(3)} = \{U_1, U_6, U_5\}, S_4^{(4)} = \{U_2, U_6, U_5\}.$ &\\
		\hline 
		$U_5$&$ S_5^{(1)} =  \{U_2, U_4\}, S_5^{(2)} =  \{U_3, U_4\}.$&$2$\\
		\hline
		$U_6$&$ S_6^{(1)} =  \{U_3, U_1\}, S_6^{(2)} =  \{U_4, U_1\}.$      & $2$   \\
		\hline
	\end{tabular}
	\label{example table}
\end{table}

%We are now in a position to describe the capacity of our heterogeneous DSS.  Using the information flow graph, the capacity of homogeneous DSS with symmetric repair (calculated in \cite{capacity}) is 
%\begin{equation}
%	\mathcal{C}(\alpha, \gamma)=\sum_{i=1}^k\min\{\alpha,(d-i+1)\frac{\gamma}{d}\}.
%\end{equation} 
%Now we can define regenerating codes for the ($n,k$) heterogeneous DSS as considered in this paper.
Similar to the parameters of the regenerating codes for homogeneous systems \cite{5550492}, we provide the parameters of Heterogeneous Regenerating codes in the next remark.
\begin{remark}
For a ($n,k$) heterogeneous DSS, regenerating codes over a field $\F_q$ are described by the parameters $[n,\ k,\ \overline{d},\overline{\alpha}, \beta, B]$, where $B$ is the file size, $\beta$ is the repair traffic, $\overline{d}$ = $[d_{1,i}=d_i]_{1\times n}$ and $\overline{\alpha}$ = $[\alpha_{1,i}=\alpha_i]_{1\times n}$ are one dimensional arrays of repair degree $d_i$ and node storage capacity $\alpha_i$ for node $U_i$ indexed with $i=1,2,\ldots, n$.
\end{remark} 

%In our ($n,k$) heterogeneous DSS, single node failure can be recovered by some surviving set as specific set of the $d_i$ helper nodes. Thus a typical information flow graph $G(\mathcal{V},\mathcal{E})$ (graphical representation of $(n, k)$ DSS) is shown in Figure \ref{Use in lemma 1}. A pair of graph nodes $In_i$ and $Out_i (1\leq i\leq n)$ in $G(\mathcal{V},\mathcal{E})$, represents the storage node $U_i$. Here node $``s"$ is the source of whole file. If $\alpha _i$ is the storage capacity of node $U_i$ then the weight of directed edge $(In_i, Out_i)$ is $\alpha _i$ in flow graph $G(\mathcal{V},\mathcal{E})$ because the node $U_i$ can flow maximum amount of information $\alpha_i$ across the graph $G(\mathcal{V},\mathcal{E})$. If a node $U_i$ ($i.e.$ node pair $(In_i, Out_i)$) fails then each helper nodes of any one of the surviving set $S_i^{(\ell)}$ for storage node $U_i$, will download $\beta$ packets and generate a new storage node $U_i'$ ($i.e.$ new node pair ($In_i', Out_i'$)). Now in order to calculate the maximum amount of packets that can be delivered to data collector ($t$) by contacting any $k$ nodes (for $G(\mathcal{V},\mathcal{E})$ any $k$ number of `out nodes' called `$Out_i$'), one has to compute the min-cut of the information flow graph $G(\mathcal{V},\mathcal{E})$. Also to compute the cut in $G(\mathcal{V},\mathcal{E})$ one requires a specific sequence of surviving sets picked up randomly one from each node. 
Note that multi-node failure can be assumed as a sequence of single node failure within a small time interval. %To repair a sequence of single node failure, sequence of respective surviving sets is needed.
So the sequence of surviving sets are needed to repair such multi node failure. Formally, the surviving sequence can be defined as follows.
 %For a $(n, k)$ heterogeneous DSS, \textit{surviving sequence} $\left\langle S_{f_j}^{(\ell_{f_j})}\right\rangle_{j=1}^n$ is a sequence of surviving sets picked up randomly one for each node $U_{f_j}$, where $f_j$ is some permutation on set $\left\lbrace 1,2,..., n\right\rbrace$ and $\ell_{f_j}\in\left\lbrace 1,2,..., \tau_{f_j}\right\rbrace $.
\begin{definition} (Surviving Sequence): For a $(n, k)$ heterogeneous DSS, surviving sequence $\left\langle S_{f_j}^{(\ell_{f_j})}\right\rangle_{j=1}^n$ is a sequence of surviving sets picked up randomly one for each node $U_{f_j}$, where $f_j$ is some permutation on set $\left\lbrace 1,2,..., n\right\rbrace$ and $\ell_{f_j}\in\left\lbrace 1,2,..., \tau_{f_j}\right\rbrace $.
\end{definition}
%For example, in $(6, 2)$ heterogeneous DSS as shown in Figure \ref{DSS example}, one of the
For some particular surviving sets, a possible surviving sequence $\left\langle S_{f_j}^{(\ell_{f_j})}\right\rangle_{j=1}^{6}$ is $\left\langle S^{(1)}_5, S^{(1)}_3,  S^{(2)}_4, S^{(2)}_6, S^{(1)}_1, S^{(2)}_2\right\rangle$ associated with failed nodes $U_5$, $U_3$, $U_4$, $U_6$, $U_1$ and $U_2$ in ($6,2$) heterogeneous DSS (see Table \ref{example table}). %Another possible surviving sequence could be $\left\langle S^{(1)}_2, S^{(1)}_3, S^{(2)}_4, S^{(2)}_6, S^{(2)}_1, S^{(2)}_5\right\rangle$.
In \cite{DBLP:journals/corr/BenerjeeG14}, heterogeneous DSS is mapped with acyclic directed graph called information flow graph. Analyzing \textit{min-cut} of the information flow graph, a fundamental bound on file size $B$ is computed for such ($n,k$) heterogeneous DSS. The bound is described in following theorem. %Hence, the necessary condition for file size $B$ as calculated in \cite{DBLP:journals/corr/BenerjeeG14}, is 

\begin{theorem}[Fundamental Bound]
	For a $(n, k)$ heterogeneous DSS, the file size $B$ must satisfy the following inequality
	
	\begin{equation*}
	B\leq \min_{\left\langle S_{f_j}^{(\ell_{f_j})}\right\rangle_{j=1}^n\in\mathscr{T}}\left\{\sum\limits_{j=1}^{k}\min \left\{\alpha_{f_j}, \left|S_{f_j}^{(\ell_{f_j})}\backslash \left(\bigcup_{\lambda=0}^{j-1}\{U_{f_{\lambda}}\}\right)\right|\beta \right\}\right\},
	\end{equation*}
	%\begin{equation}
	%\begin{split}
	%& B\leq \\ & \min_{\left\langle S_{f_j}^{(\ell_{f_j})}\right\rangle_{j=1}^n\in\mathscr{T}}\left\{\sum\limits_{j=1}^{k}\min \left\{\alpha_{f_j}, \left|S_{f_j}^{(\ell_{f_j})}\backslash \left(\bigcup_{\lambda=0}^{j-1}\{U_{f_{\lambda}}\}\right)\right|\beta \right\}\right\},
	%\end{split}
	%\label{condition for B 1}
	%\end{equation}
	where $\{U_{f_0}\}=\phi,\ 0\leq \lambda < j\leq k$, $S_{f_j}^{(\ell_{f_j})}\in\left\langle S_{f_j}^{(\ell_{f_j})}\right\rangle_{j=1}^n$, $\mathscr{T}$ is the set of all surviving sequences $\left\langle  S_{f_j}^{(\ell_{f_j})} \right\rangle_{j=1}^n$ with length $n$ and $\ell_{f_j}\in\left\lbrace 1,2,...,\tau_{f_j}\right\rbrace$.
	\label{condition for B 1}
\end{theorem}

In \cite{DBLP:journals/corr/BenerjeeG14}, it is shown that there exist code which achieves the fundamental bound for such ($n,k$) heterogeneous DSS. Hence, one can get the optimal codes by reducing parameters which meets the fundamental bound. In the next section, parameters for the optimal codes are computed by minimizing node storage capacity and repair bandwidth.

%As per the required DSS optimization, the conditions (by minimizing node storage capacity and then repair bandwidth) which achieves the Bound \ref{condition for B 1}, is computed in the next section.

%%%%%%%%%%%%%%%%%%%%
\section{Conditions for optimality}
%In this section we will discuss about family of codes which achieves the Bound \ref{condition for B 1}. 

Consider a $(n,k)$ heterogeneous DSS with $\tau_i$ number of surviving sets $S_i^{(\ell_i)}$ and repair degree $\left| S_i^{(\ell_i)}\right|=d_i$ ($\ell_i = 1,2,\ldots,\tau_i$ and $i = 1,2,\ldots,n$). If $\alpha_i>\left| S_i^{(\ell_i)}\right|\beta$ then the failed node $U_i$ can not be repaired so $\alpha_i\leq\left| S_i^{(\ell_i)}\right|\beta$ for each $i$ and $\ell_i$. For optimality, %at $d_i$ = $\min\limits_{1\leq\ell_i\leq\tau_i}\left| S_i^{(\ell_i)}\right|$,
 $\alpha_i=d_i\beta$. Hence for constant repair traffic $\beta$, node storage capacity $\alpha_i$ and repair degree $d_i$ are proportional to each other. Consider $c_i\in(0,1)\subset\RR$ such that $\sum\limits_{i=1}^{n}c_i = 1$ and $\dfrac{c_i}{\alpha_i}=\dfrac{c_j}{\alpha_j}$ for $1\leq i<j\leq n$. %and $i,j\in\ZZ$. 
Hence, %	c_i=\alpha_i\left(\sum\limits_{i=1}^{n}\alpha_i\right)^{-1}.
$\alpha_i$ = $c_i\sum\limits_{i=1}^{n}\alpha_i$ = $c_i\alpha^*$ ($i=1,2,\ldots,n$). %=\frac{d_i}{\sum\limits_{i=1}^{n}d_i}.
%where without loss of generalization, one can assume $\alpha_1\leq\alpha_2\leq...\leq\alpha_n$.
%To store a file in the $(n,k)$ heterogeneous DSS, the file is divided into $B$ distinct massage packets. The symbols for the packets are from the field $\F_q$. Encode the $B$ packets using linear code and distribute the packets on $n$ distinct nodes such that an arbitrary node $U_i$ $(i=1,2,...,n)$ has $\alpha_i$ number of packets and an arbitrary failed node $U_i$ can be repaired by helper nodes of some surviving set $S_i^{(\ell_i)}$.
So, the parameters $\alpha_i$ and $c_i$ are proportional to each other. %$i.e.$ $\exists$ $\alpha^*$ $s.t.$ $\alpha_i = c_i\alpha^*$ ($i=1,2,...,n$). 
Again, $k$ is reconstruct degree so, $B\leq \sum_{i\in\mathcal{K}}\alpha_i =\sum_{i\in\mathcal{K}}c_i\alpha^*$ for any arbitrary set $\mathcal{K}\subset\{1,2,...,n\}$ such that $|\mathcal{K}|=k$. Hence, $B\leq \sum_{i=1}^kc_i\alpha^*$ for $c_1\leq c_2\leq ...\leq c_n$. For optimum case, one can reduce $\alpha^*$ up to $\alpha^*_{min}$ such that
\begin{equation}
B = \sum_{i=1}^{k}c_i\alpha^*_{min}\Rightarrow\alpha^*_{min}=B\left( \sum_{j=1}^{k}c_j\right) ^{-1},
\label{c}
\end{equation} %where $c_1\leq c_2\leq...\leq c_n$.

%Since the Bound \ref{condition for B 1} is tight, a heterogeneous DSS having the minimized parameters $\alpha_i$ and $\beta$, will hold the Bound \ref{condition for B 1} with equality. Because of the parameter $\alpha_i$ is proportional to $c_i$ ($\alpha_i = c_i\alpha^*,\;\forall \;i\in\{1,2,...,n\}$) and any $k$ nodes are able to reconstruct the whole file, one can reduce node storage by reducing proportional factor $\alpha^*$. Again, any $k$ nodes are able to reconstruct the file so $\sum\limits_{i=1}^{k}\alpha_i$ = $B$. Hence the minimum possible proportional factor \[\alpha^*_{min}=\frac{B}{\sum\limits_{j=1}^{k}c_j},\] where $c_1\leq c_2\leq...\leq c_n$.

Similarly for a fixed proportional factor $\alpha_{min}^*$, one can minimize the repair traffic $\beta$ such that Bound \ref{condition for B 1} holds with equality. For a specific surviving sequence $\left\langle S_{f_j}^{(\ell_{f_j})}\right\rangle_{j=1}^n$ with sufficient large repair traffic $\beta$, the inequality $\alpha_{f_m}\leq\left|S_{f_m}^{(\ell_{f_m})}\backslash \left(\bigcup\limits_{\lambda=0}^{m-1}\{U_{f_{\lambda}}\}\right)\right|\beta$ holds for each  $m=1,2,\ldots,k$. %Hence, $\alpha_{f_m} \left|S_{f_m}^{(\ell_{f_m})}\backslash \left(\bigcup\limits_{\lambda=0}^{m-1}\{U_{f_{\lambda}}\}\right)\right|^{-1}\leq\beta$.
%One can reduce repair traffic $\beta$ such that at least one of the inequality among $k$ number of inequalities (as given in  \ref{beta}) reduces up to equality. 
If we choose $\beta$ = $\beta_{min}$ such that 
\begin{equation}
\beta_{min} = \max\limits_{\left\langle S_{f_j}^{(\ell_{f_j})}\right\rangle_{j=1}^n\in\mathscr{T}}\left\lbrace\max\limits_{1\leq m\leq k}\left\lbrace \alpha_{f_m} \left|S_{f_m}^{(\ell_{f_m})}\backslash \left(\bigcup\limits_{\lambda=0}^{m-1}\{U_{f_{\lambda}}\}\right)\right|^{-1}\right\rbrace \right\rbrace 
\label{beta min}
\end{equation}

%So, for a given surviving sequence $\left\langle S_{f_j}^{(\ell_{f_j})}\right\rangle_{j=1}^n$, if %there exist  $\beta$ = $\beta\left( \left\langle S_{f_j}^{(\ell_{f_j})}\right\rangle_{j=1}^n\right)$ such that
%$\beta\left( \left\langle S_{f_j}^{(\ell_{f_j})}\right\rangle_{j=1}^n\right) =\max\limits_{1\leq m\leq k}\left\lbrace \alpha_{f_m} \left|S_{f_m}^{(\ell_{f_m})}\backslash \left(\bigcup\limits_{\lambda=0}^{m-1}\{U_{f_{\lambda}}\}\right)\right|^{-1}\right\rbrace$
%then $\beta\left( \left\langle S_{f_j}^{(\ell_{f_j})}\right\rangle_{j=1}^n\right)$ is the minimum value of repair traffic $\beta$ for the specific surviving sequence that meets the Fundamental Bound \ref{condition for B 1}. Now, if 
%\begin{equation}
%\beta_{min} = \max\left\lbrace \beta\left( \left\langle S_{f_j}^{(\ell_{f_j})}\right\rangle_{j=1}^n\right) :\left\langle S_{f_j}^{(\ell_{f_j})}\right\rangle_{j=1}^n\in\mathscr{T}\right\rbrace 
%\label{beta min}
%\end{equation}
then $\beta_{min}$ is the minimum value of repair traffic $\beta$ which ensures  $\left|S_{f_m}^{(\ell_{f_m})}\backslash \left(\bigcup\limits_{\lambda=0}^{m-1}\{U_{f_{\lambda}}\}\right)\right|\beta_{min}\geq$  $\alpha_{f_m}$ for each $f_m$ of an arbitrary surviving sequence. 

Formally the results can be summarized by the following theorem.
\begin{theorem}
	Consider a ($n,k$) heterogeneous DSS with given surviving sets $S_i^{(\ell_i)}$ ($i=1,2,...,n$; $\ell_i=1,2,...,\tau_i$ for some $\tau_i\in\ZZ$). %$c_i\in(0,1)\subset\RR$ such that $\sum\limits_{i=1}^nc_i=1$, $c_i\leq c_j$ and $\frac{c_i}{\alpha_i}$=$\frac{c_i}{\alpha_i}$ ($1\leq i<j\leq n$; $i,j\in\ZZ$), 
	A family of codes with %$B$ = $\sum\limits_{j=1}^k\alpha_j$,
	 $\alpha_i$ = $c_i\alpha_{min}$ = $d_i\beta_{min}$ and $\beta$= $\beta_{min}$, achieves the Fundamental Bound \ref{condition for B 1}, where $\alpha_{min}$ and $\beta_{min}$ can be calculated by Equations (\ref{c}) and (\ref{beta min}).
\end{theorem}
%\begin{proof}
%	It is easy to see that the bound in Inequality (\ref{condition for B 1}) holds for an arbitrary surviving sequence $\left\langle S_{f_j}^{(\ell_{f_j})}\right\rangle_{j=1}^n\in\mathscr{T}$ with $\alpha_{f_m}\leq\left|S_{f_m}^{(\ell_{f_m})}\backslash \left(\bigcup\limits_{\lambda=0}^{m-1}\{U_{f_{\lambda}}\}\right)\right|\beta$ (each $m\in\{1,2,...,k\}$) and $B = \sum\limits_{m=1}^k\alpha_m\;(\mbox{for }\alpha_1\leq\alpha_2\leq...\leq\alpha_n)$.

	%For an arbitrary surviving sequence $\left\langle S_{f_j}^{(\ell_{f_j})}\right\rangle_{j=1}^n\in\mathscr{T}$, $\alpha_{f_m}\leq\left|S_{f_m}^{(\ell_{f_m})}\backslash \left(\bigcup\limits_{\lambda=0}^{m-1}\{U_{f_{\lambda}}\}\right)\right|\beta$ (each $m\in\{1,2,...,k\}$). Hence, 
	%\begin{equation}
%	\begin{split}
%	 \min_{\left\langle  S^{(\ell_{f_j})}_{f_j}\right\rangle_{j=1}^n\in\mathscr{T}}& \left\{\sum\limits_{j=1}^{k}\min \left\{\alpha_{f_j}, \left|S^{(\ell_{f_j})}_{f_j}\backslash \left(\bigcup_{\lambda=0}^{j-1}\{U_{f_{\lambda}}\}\right)\right|\beta \right\}\right\}\\
%	 & = \min_{\left\langle  S^{(\ell_{f_j})}_{f_j}\right\rangle_{j=1}^n\in\mathscr{T}} \left\{\sum\limits_{j=1}^{k}\alpha_{f_j}\right\}\\
%	 & = \sum\limits_{m=1}^k\alpha_m\;(\mbox{for }\alpha_1\leq\alpha_2\leq...\leq\alpha_n)\\
%	 & = B.
%	 \end{split}
%	 \end{equation}
%\end{proof}
%Since $\sum\limits_{i=1}^{k}\alpha_i$ = $B$ (for $\alpha_1\leq\alpha_2\leq...\leq\alpha_n$), $\beta$ can be reduced up to the associated node storage capacity $\alpha_{f_{j^*}}$ (for some $\;j^*\in\{1,2,...,k\}$), where $f_{j^*}$ is some permutation on $\{1,2,...,k\}$. 
Next section presents a construction of an optimal family of regenerating code based on graph. 
%Mathematically, for some $j^*\in\{1,2,...,k\}$, one can have

%\begin{equation}
%\begin{split}
% & \min\limits_{\left\langle S_{f_m}^{(\ell_{f_m})}\right\rangle_{m=1}^k                 }\min\left\lbrace \left|S_{f_m}^{(\ell_{f_m})}\backslash \left(\bigcup\limits_{\lambda=0}^{j-1}\{U_{f_{\lambda}}\}\right)\right|: j=1,2,...,k\right\rbrace\\  &=\left|S_{f_{j^*}}^{(\ell_{f_{j^*}})}\backslash \left(\bigcup\limits_{\lambda=0}^{j^*-1}\{U_{f_{\lambda}}\}\right)\right|
% \end{split}
%\end{equation}
%one can find $\alpha_{j^*}$ = $\left|\eta_{j^*}\backslash \left(\bigcup\limits_{\lambda=0}^{j^*-1}\{U_{\lambda}\}\right)\right|\beta_{min}$. Hence \[\beta_{min}=\dfrac{\left( \frac{B}{\sum\limits_{j=1}^{k}c_j}\;c_{j^*}\right) }{\left|\eta_{j^*}\backslash \left(\bigcup\limits_{\lambda=0}^{j^*-1}\{U_{\lambda}\}\right)\right|}\]
%\begin{remark} In the particular ($n,k$) heterogeneous DSS, $ \theta$ = $\sum\limits_{i=1}^{n}\alpha_i$
% and $B$ = $\sum\limits_{j=1}^{k}\alpha_j$.
% \end{remark}
 
% If a file with size $B$ is stored in a ($n,k$) heterogeneous DSS with $c_i$ $(1\leq i\leq n)$, then $\alpha_i=\frac{B\;c_i}{\sum\limits_{j=1}^{k}c_j}$, $B$ = $\sum\limits_{j=1}^{k}\alpha_j$ and $\beta=\dfrac{ B\;c_{j^*}}{\left|\eta_{j^*}\backslash \left(\bigcup\limits_{\lambda=0}^{j^*-1}\{U_{\lambda}\}\right)\right|\sum\limits_{j=1}^{k}c_j}$. %and $d_i$ = $\left|\eta_{j^*}\backslash \left(\bigcup\limits_{\lambda=0}^{j^*-1}\{U_{\lambda}\}\right)\right|\frac{c_i}{c_{j^*}}$, where $\alpha_i=d_i\beta$.

%%%%%%%%%%%%%%%%%%%%
\section{Family of Optimal Codes}
Graph $\mathscr{H}(\mathcal{V},\mathcal{E})$ is representation of vertex set $\mathcal{V}$ and edge set $\mathcal{E}$ such that $\mathcal{E}$ = $\{E:|E|=2\mbox{ and }E\subset\mathcal{V}\}$. For a graph $\mathscr{H}(\mathcal{V},\mathcal{E})$, degree of vertex $v_i\in\mathcal{V}$ (denoted by $deg(v_i)$) is the total number of edges $E\in\mathcal{E}$ such that $v_i\in E$. An edge $E\in\mathcal{E}$ is called loop if $E=\{v_i,v_i\}$ for some $i\in\{1,2,\ldots,n\}$. Similarly, two edges $E_l$, $E_m\in\mathcal{E}$ are called parallel edges if $E_l$ = $E_m$ ($l\neq m$). A graph $\mathscr{H}(\mathcal{V},\mathcal{E})$ is called simple if the graph does not have loop or parallel edges. Two vertices $v_i$ and $v_j$ are adjacent if $\exists\; E\in\mathcal{E}$ $s.t.$ $E=\{v_i,v_j\}$. For two arbitrary vertices $v_i$ and $v_j$, if there exist a sequence of vertices having $v_i$ and $v_j$ such that arbitrary two consecutive vertices in the sequence are adjacent vertices then the particular graph $\mathscr{H}(\mathcal{V},\mathcal{E})$ is called connected graph.

For vertex set $\mathcal{V}$ and edge set $\mathcal{E}$, let $\mathscr{H}(\mathcal{V},\mathcal{E})$ be a simple connected graph such that % with $deg(v_i)\leq deg(v_j)$ (for $1\leq i<j\leq n$) such that
\begin{enumerate}
	\item $2\leq deg(v_i)\leq deg(v_j)$, for $1\leq i<j\leq n$ and
	\item $\{v_i,v_j\}$ $\notin \mathcal{E}$, for each $i,j$ ($1\leq i<j\leq k$).
\end{enumerate}

%where $deg(v_i)$ is degree of vertex $v_i\in\mathcal{V}$ ($i=1,2,...,n$). 
Distribute all $\sum_{i=1}^n\alpha_i$ encoded packets among all nodes $U_i$ ($i=1,2,...,n$) $i.e$ $\alpha_i$ = $|U_i|$ = $deg(v_i)$. The particular distribution can be done into following two parts. 
\begin{enumerate}
	\item  Randomly distribute $B$ encoded packets among $k$ nodes say $U_l$ ($l=1,2,\ldots,k$).
	\item Distribute the remaining $\sum_{p=k+1}^n\alpha_p$ packets on left $n-k$ nodes $U_j$ ($j=k+1,k+2,\ldots,n$) such that the particular packet stored on node $U_j$ is the linear combination of packets stored in nodes $U_m$, where $m$ is a positive integer %is associated with vertex $v_j$ 
	such that $\{v_j,v_m\}\in\mathcal{E}$ and $m<j$. The linear combinations are taken such a way that any two packets stored in same node are linearly independent.
\end{enumerate}

 Observe that $S_i^{(1)}\subseteq\left\lbrace U_m: \forall\ \{v_i,v_m\}\in\mathcal{E}\right\rbrace$. Consider a simple case $\beta$ = $1$ and $S_i^{(1)}$ = $\left\lbrace U_m: \{v_i,v_m\}\in\mathcal{E}\right\rbrace$. In this case, $deg(v_i)$ = $\alpha_i$ = $d_i$ = $|S_i^{(1)}|$, $\tau_i$ = $1$  %Since $deg(v_i)$ = $\left| S_{\Psi(i)}^{(1)}\right|$ = $d_{\Psi(i)}$ = $\alpha_{\Psi(i)}\beta^{-1}$ (for optimum case $\alpha_{\Psi(i)} = d_{\Psi(i)}\beta$),
 and $B=\sum_{i=1}^kdeg(v_i)$. Observe that $deg(v_i)$ and $c_i$ ($1\leq i\leq n$) are proportional to each other. %For optimum bandwidth, if $\beta=1$ then $\alpha_{\Psi(i)}$ = $d_{\Psi(i)}$. Hence $\alpha_{\Psi(i)}$ = $deg(v_i)$ and $B=\sum_{i=1}^kdeg(v_i)$.

For example, %need to store among remaining nodes $U_{\Psi(j)}$ ($j=k+1,k+2,...,n$) such that $|U_{\Psi(j)}|$ = $deg(v_j)$. The particular distribution is done , are the linear combinations of packets stored in nodes $U_m$, where index  $m\in\{1,2,\ldots,j-1\}$ is associated with vertex $v_j$ such that $\{v_j,v_m\}\in\mathcal{E}$. For some particular linear combinations, one can observe that $\left\lbrace U_m: \{v_j,v_m\}\in\mathcal{E}\right\rbrace=S_i^{(1)}$. Since $deg(v_i)$ = $\left| S_i^{(1)}\right|$ = $d_i$ = $\alpha_i\beta^{-1}$ (for optimum case $\alpha_i = d_i\beta$), $deg(v_i)$ is proportional to $c_i$ ($1\leq i\leq n$). For optimum bandwidth, if $\beta=1$ then $\alpha_i$ = $d_i$. Hence $\alpha_i$ = $deg(v_i)$ and $B=\sum_{i=1}^kdeg(v_i)$.
%In this graph $\mathscr{H}(\mathcal{V},\mathcal{E})$, each vertex $v_i\in\mathcal{V}$ is associated with node $U_i$ in $(n,k)$ heterogeneous DSS. Clearly $|\mathcal{V}|=n$. %and degree of vertex $v_i\in\mathcal{V}$ is $deg(v_i)$ ($\forall i,\ i\in\{1,2,...,n\}$).
%One can connect $deg(v_i)$ with node storage capacity $\alpha_i$.  %WLOG, one can assume $deg(v_i)\leq deg(v_j)$ for each $(1\leq i<j\leq n)$.
%First $B$ encoded packets are distributed randomly among all nodes $v_i$ ($i=1,2,...,k$). The remaining $\sum_{i=k+1}^n\alpha_i$ packets are stored among remaining nodes $U_j$ ($j\in\{k+1,k+2,...,n\}$). The particular distribution is done such a smart way that $\alpha_j$ number of packets stored in node $U_j$, are the linear combinations of packets stored in nodes $U_m$, where index  $m\in\{1,2,\ldots,j-1\}$ is associated with vertex $v_j$ such that $\{v_j,v_m\}\in\mathcal{E}$. For some particular linear combinations, one can observe that $\left\lbrace U_m: \{v_j,v_m\}\in\mathcal{E}\right\rbrace=S_i^{(1)}$. Since $deg(v_i)$ = $\left| S_i^{(1)}\right|$ = $d_i$ = $\alpha_i\beta^{-1}$ (for optimum case $\alpha_i = d_i\beta$), $deg(v_i)$ is proportional to $c_i$ ($1\leq i\leq n$). For optimum bandwidth, if $\beta=1$ then $\alpha_i$ = $d_i$. Hence $\alpha_i$ = $deg(v_i)$ and $B=\sum_{i=1}^kdeg(v_i)$.
  %In this construction observe, $k<n=\left| \mathcal{V}\right|$, $\beta=1$ $unit$, $\alpha_i$ = $d_i$ = $deg(v_i)$ and $B=\sum_{i=1}^kdeg(v_i)$.
	 consider a simple connected graph $\mathscr{H}(\mathcal{V},\mathcal{E})$ with degree sequence $2$, $2$, $3$, $3$ as shown in Figure \ref{example}. Suppose, $4$ $\left(=B=\sum_{i=1}^{2}deg(v_i)\right)$ packets are encoded into $10\ \left(=\sum_{i=1}^4deg(v_i)\right)$ packets such that $y_i$ = $x_i$ for $i=1,2,3,4$ and $y_j$ = $f_{j-4}(x_1,x_2,x_3,x_4)$ for $j=5,6,7,8,9,10$. In particular, $f_{j-4}(x_1,x_2,x_3,x_4)$ is linear function of $x_1,x_2,x_3$ and $x_4$ on $\FF_q$. 
	 	 % Define block styles
	 	 \tikzstyle{line} = [draw, -]
	 	 \tikzstyle{cloud} = [draw, circle,fill=red!20, node distance=2cm, minimum height=2em]
	   	 \tikzstyle{block 1} = [rectangle, draw, fill=red!20, text width=4em, text centered, rounded corners, minimum height=1em, node distance=3.5cm]
	 	 \tikzstyle{block 2} = [rectangle, draw, fill=red!20, text width=4em, text centered, rounded corners, minimum height=1em, node distance=0.7cm]
	 	 \tikzstyle{block 3} = [rectangle, draw, fill=red!20, text width=5em, text centered, rounded corners, minimum height=1em, node distance=0.7cm]
	 	  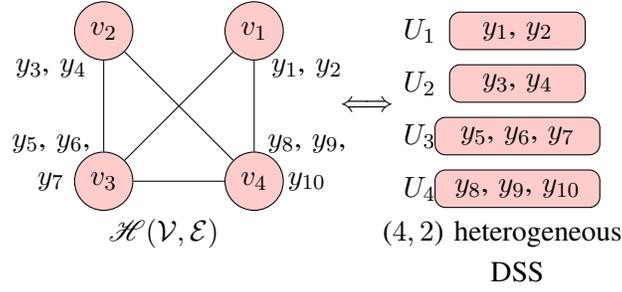
\begin{figure}\center
	 	 \begin{tikzpicture}[node distance = 2cm, auto]
	 	 % Place nodes
	 	 \node [cloud] (1) {$v_1$};
	 	 \node [cloud, left of=1] (2) {$v_2$};
	 	 \node [cloud, below of=1] (identify) {$v_4$};
	 	 \node [cloud, left of=identify] (evaluate) {$v_3$};
	 	 \path [line] (1) -- (identify);
	 	 \path [line] (identify) -- (evaluate);
	 	 \path [line] (1) -- (evaluate);
	 	 \path [line] (2) -- (identify);
	 	 \path [line] (2) -- (evaluate); 
	 	 \draw (0.7,-0.5) node {$y_1$, $y_2$};
	 	 \draw (-2.7,-0.5) node {$y_3$, $y_4$};
	 	 \draw (-2.7,-1.5) node {$y_5$, $y_6$,};
	 	 \draw (-2.7,-2) node {$y_7$};
	 	 %\draw (-4.5,-3.5) node {$x_1+x_3$,};
	 	 %\draw (-5.2,-4) node {$x_1+x_2+x_3+x_4$};
	 	 \draw (0.7,-1.5) node {$y_8$, $y_9$,};
	 	 \draw (0.7,-2) node {$y_{10}$};
	 	 \draw (1.5,-1) node {$\Longleftrightarrow$};
	 	 \node [block 1, right of=1] (3) {$y_1$, $y_2$};
	 	 \node [block 2, below of=3] (4) {$y_3$, $y_4$};
	 	 \node [block 3, below of=4] (5) {$y_5$, $y_6$, $y_7$};
	 	 \node [block 3, below of=5] (6) {$y_8$, $y_9$, $y_{10}$};
	 	 \draw (2.2,0) node {$U_1$};
	 	 \draw (2.2,-0.7) node {$U_2$};
	 	 \draw (2.2,-1.4) node {$U_3$};
	 	 \draw (2.2,-2.1) node {$U_4$};
	 	 \draw (-1.2,-2.7) node {$\mathscr{H}(\mathcal{V},\mathcal{E})$};
	 	 \draw (3.3,-2.7) node {($4,2$) heterogeneous};
	 	 \draw (3.5,-3.2) node { DSS};
	 	 %\draw (-3,-2.7) node {Fig. 3 An example of construction of DSS with achieving bounds};
	 	 \end{tikzpicture}
	 	 \caption{Relation between graph $\mathscr{H}(\mathcal{V},\mathcal{E})$ and ($4,2$) heterogeneous DSS is illustrated, where $\mathcal{V}$=$\{v_1,v_2,v_3,v_4\}$ and $\mathcal{E}$ = $\left\lbrace \{v_1,v_3\}, \{v_1,v_4\},\{v_2,v_3\},\right.$ $\left.\{v_2,v_4\},\{v_3,v_4\}\right\rbrace $.}
	 	 \label{example}
	      \end{figure}
	 For $\alpha_i$ = $d_i$ = $deg(v_i)$, %such that $y_1$ = $x_1$, $y_2$ = $x_2$, $y_3$ = $x_3$, $y_4$ = $x_4$, $y_5$ = $x_5$, $y_6$ = $x_6$, $y_7$ = $x_7$, $y_8$ = $f_1(x_i;i=1,2,...,7)$, $y_9$ = $f_2(x_i;i=1,2,...,7)$ and $y_{10}$ = $f_3(x_i;i=1,2,...,7)$, where $f_j(x_i;i=1,2,...,7)$ are linear independent vectors for $j=1,2,3$. 
	 one can choose $y_5$ = $x_2$, $y_6$ = $x_2+x_4$, $y_7$ = $x_1+x_2+x_3+x_4$, $y_8$ = $x_1$, $y_9$ = $x_1+x_3$ and $y_{10}$ = $x_1+x_2+x_3+x_4$. For the respective ($4,2$) heterogeneous DSS, packets $y_j\ (j=1,2,...,10)$ are distributed among $4$ ($=n$) nodes (see Figure \ref{example}). Observe that $S_1^{(1)}$ = $\{U_3, U_4\}$, $S_2^{(1)}$ = $\{U_3, U_4\}$, $S_3^{(1)}$ = $\{U_1, U_2,  U_4\}$ and $S_4^{(1)}$ = $\{U_1, U_2, U_3\}$ and $\beta_{min}=1$. %encoded packets indexed by $1$ to $7$ are stored in nodes $U_1$, $U_2$ and $U_3$ such that $2$ packets are stored in node $U_1$, another two packets are in $U_2$ and renaming three are in node $U_3$. Since $v_3$ is adjacent to $v_1$, $v_2$ and $v_3$ so packets $y_8$, $y_9$ and $y_{10}$ are stored in node $U_4$.
	 %For a surviving sequence $\left\langle S_2^{(1)}, S_3^{(1)}, S_4^{(1)}, S_1^{(1)}\right\rangle $, $\beta\left(\left\langle S_2^{(1)}, S_3^{(1)}, S_4^{(1)},\right. \right. $ $\left. \left. S_1^{(1)}\right\rangle\right)$ = $\max\left\lbrace \dfrac{\alpha_2}{\left| S_2^{(1)}\right|},\dfrac{\alpha_3}{\left| S_3^{(1)}\backslash\{U_2\}\right|} \right\rbrace$ = $1.5$ $unit$. By some calculation one can find $\beta_{min}$ = $1.5$ $unit$. 
	 %For the surviving sequence  $\left\langle S_1^{(1)}, S_2^{(1)}, S_3^{(1)}, S_4^{(1)}\right\rangle $, $\sum\limits_{j=1}^{2}\min \left\{\alpha_{j},\right. $ $\left.  \left|S_{j}^{(\ell_{j})}\backslash \left(\bigcup_{\lambda=0}^{j-1}\{U_{\lambda}\}\right)\right|\beta \right\}$ %= $\min\{2,2\}+\min\{2,2\}$ 
 Hence, the graphical construction %\footnote{One can generalize the construction for large $\beta$ and $\tau_i$.}%\footnote{One can generalize the construction for higher values of the repair traffic and large number of surviving sets.} 
 can be summarized by the following theorem.
 \begin{theorem}
       Consider a simple connected graph $\mathscr{H}(\mathcal{V},\mathcal{E})$ with $2\leq deg(v_m)\leq deg(v_j)$ (for $1\leq m<j\leq n$) such that two arbitrary vertices from $\{v_1,v_2,\ldots,v_k\}$ are not adjacent through an edge. A ($n=\left|\mathcal{V}\right|,k=\left|\{v_1,v_2,\ldots,v_k\}\right|$) heterogeneous DSS associated with the graph $\mathscr{H}(\mathcal{V},\mathcal{E})$, achieves the Fundamental Bound \ref{condition for B 1}, where the node storage capacity $\alpha_i$ and repair degree $d_i$ are each $deg(v_i)$ and repair traffic $\beta$ is $1$.
 \end{theorem}
%\begin{remark} One can generalize the construction for higher values of the repair traffic and large number of surviving sets. \end{remark}

%%%%%%%%%%%%%%%%%%%%%%%%%%%%%%%%%%%%%
\section{Conclusion}
Motivated by the real world applications we considered heterogeneous DSSs with dynamic repair degree and node storage capacity.
A graphical construction framework is used to construct a family of optimal regenerating codes meeting heterogeneous DSS fundamental bound on file size.
\small
%\footnotesize
\bibliographystyle{IEEEtran}
\bibliography{cloud}

\end{document}